\documentclass[12pt]{article}

\usepackage[margin=1in]{geometry}
\usepackage{setspace}
\usepackage{amsmath,amssymb,amsthm,mathtools}
\usepackage{bm}
\usepackage{graphicx}
\usepackage{color}
\usepackage{hyperref}

\usepackage{algorithm}
\usepackage{algpseudocode}
\usepackage{bbm}
\usepackage{dcolumn}
\newcolumntype{L}{D{.}{.}{2,5}}

\usepackage{parskip}
\usepackage{dcolumn}
\newcolumntype{L}{D{.}{.}{2,5}}

\usepackage[round]{natbib}

\begin{document}

\begin{center}

{\LARGE An Observational Study of the Effect of Nike Vaporfly Shoes on
Marathon Performance}

\vspace{16pt}

{Joseph Guinness, Debasmita Bhattacharya, Jenny Chen, Max Chen, Angela Loh }
\vspace{8pt}

\textit{Cornell University}
\vspace{16pt}

\textbf{Abstract}
\end{center}

We collected marathon performance data from a systematic sample of
elite and sub-elite athletes over the period 2015 to 2019, then
searched the internet for publicly-available photographs of these
performances, identifying whether the Nike Vaporfly shoes were worn or
not in each performance. Controlling for athlete ability and race
difficulty, we estimated the effect on marathon times of wearing the
Vaporfly shoes. Assuming that the effect of Vaporfly shoes is
additive, we estimate that the Vaporfly shoes improve men's times
between 2.0 and 3.9 minutes, while they improve women's times
between 0.8 and 3.5 minutes. Assuming that the effect of Vaporfly
shoes is multiplicative, we estimate that they improve men's times
between 1.4 and 2.8 percent and women's performances between 0.6 and
2.2 percent. The improvements are in comparison to the shoe the
athlete was wearing before switching to Vaporfly shoes, and represents
an expected improvement rather than a guaranteed improvement.

\noindent

\section{Introduction}\label{introduction}

There is a growing consensus that Nike Corporation's new line of
marathon racing shoes, which are commonly referred to as Vaporflys,
provide a significant performance advantage to athletes who wear
them. While several different versions of the shoes have appeared in
races, including the Vaporfly 4\%, the Vaporfly Next\%, the Alphafly,
and several prototype shoes, each iteration of the shoes has in common
a carbon fiber plate stacked inside of a highly responsive foam sole.

Several research studies have investigated the magnitude of the
Vaporfly performance benefit. \cite{hoogkamer2018comparison} and
\cite{barnes2019randomized} tested highly trained distance runners in
laboratory studies, measuring various biomechanical and physiological
variables while subjects wore Vaporflys and several other shoes in
trial runs on a treadmill. Although the measured benefits varied
somewhat from athlete to athlete, both studies found a roughly 4\%
average reduction in energy expenditures while wearing Vaporflys, in
comparison to other popular racing shoes such as the Adidas Adizero
Adios Boost line of racing shoes, and Nike Zoom Matumbo track
spikes.

The Upshot, a division of The New York Times, collected data from
actual marathon performances recorded on Strava, a popular running log
and GPS tracking website. Their study included hundreds of thousands
of marathon performances, and dozens of different shoes. The Upshot
found that the Vaporflys imparted a 4 to 5\% advantage in finishing
time over an average shoe and a 1.5 to 2.5\% advantage over the
second-best shoe \citep{upshot1,upshot2}. A study published by Wired
Magazine found that a sample of runners in the 2017 New York City
Marathon were more likely to run the second half of the race faster
than the first if they were wearing Vaporflys \citep{wired1}.

Our study is most similar to the Upshot study in that we analyze data
from marathon performances and compare people's performances with and
without the Vaporflys. However, our study differs in a few
ways. First, instead of relying on a convenient sample of athletes who
upload their data to Strava, we take an exhaustive sample of athletes who
met a minimum performance standard at one of 22 of the largest
marathon venues in 2015 and 2016 in the US and Canada. Second, instead
of relying on self-reported shoe data, we searched the internet for
photos of races and visually identified the shoes that runners
wore. Third, we focus only on athletes who performed at an elite
level before the Vaporflys were released to the public. Thus, we are
only considering accomplished runners with marathon experience who,
most likely, settled on a suitable shoe before the
Vaporflys were released. These runners are also those most likely to
be affected by shoe regulations because many of them compete in
national Olympic qualifying races subject to regulations.

\section{Study Design}\label{study_design}

We selected athletes who recorded a sufficiently fast marathon
time---men under 2:24 and women under 2:45---at a collection of 22
distinct marathon venues in 2015 or 2016, including the 2016 U.S.\
Olympic Marathon Trials, which were contested in Los Angeles in
February of 2016. The list of marathons is included in the
Appendix. This resulted in a sample of 270 distinct women and 308
distinct men after matching names and our best effort to correct
alternate spellings of names. We recorded these athletes' performances
in the same 22 marathon venues over the period 2015 to 2019, and
searched publicly available online photographs, manually identifying
whether or not each athlete was wearing a Nike Vaporfly shoe by visual
inspection. All marathon times were downloaded from the website
www.marathonguide.com.

Our criteria for inclusion in the study were meant to satisfy certain
objectives. First, we wanted to study elite and sub-elite athletes,
since shoe regulations are motivated by performance
advantages for athletes in this group.  Second, we wanted to study
athletes who had achieved success in the marathon before the Nike
Vaporfly shoes had been released to the public. This ensures that
inclusion in the study is unrelated to whether an athlete was wearing
the shoes in the race where they qualified for inclusion in the
study. This is important because, if any shoe effect exists, the
magnitude of the effect may differ among different athletes. If we
were to use performances potentially aided by the shoes to select the
athletes, that might have biased our sample towards athletes who
benefit most from the shoes.

To identify shoes worn by the runners, we used photos posted on public
websites such as marathonfoto.com, marathon-photos.com,
sportphoto.com, and flashframe.io. We also collected photographs from
social media sites such as facebook.com and instagram.com. We assumed
that Vaporfly shoes were not worn in 2015 or 2016 by any runners
except for a few that were reported to have worn prototypes in the
2016 US Olympic Trials Marathon. Identification of shoes via photos is
a manual process that is subject to error. We have made all of our
shoe identifications publicly available at
\url{https://github.com/joeguinness/vaporfly} and will update this
paper with new data if we are made aware of any errors in shoe
identification.  We identified the shoes worn in 840 of 880 (95.5\%)
men's performances in our dataset and in 778 of 810 (96.0\%) women's
performances.

\section{Data Exploration}\label{data_exploration}

\begin{figure}
\centering
\includegraphics[width=\textwidth]{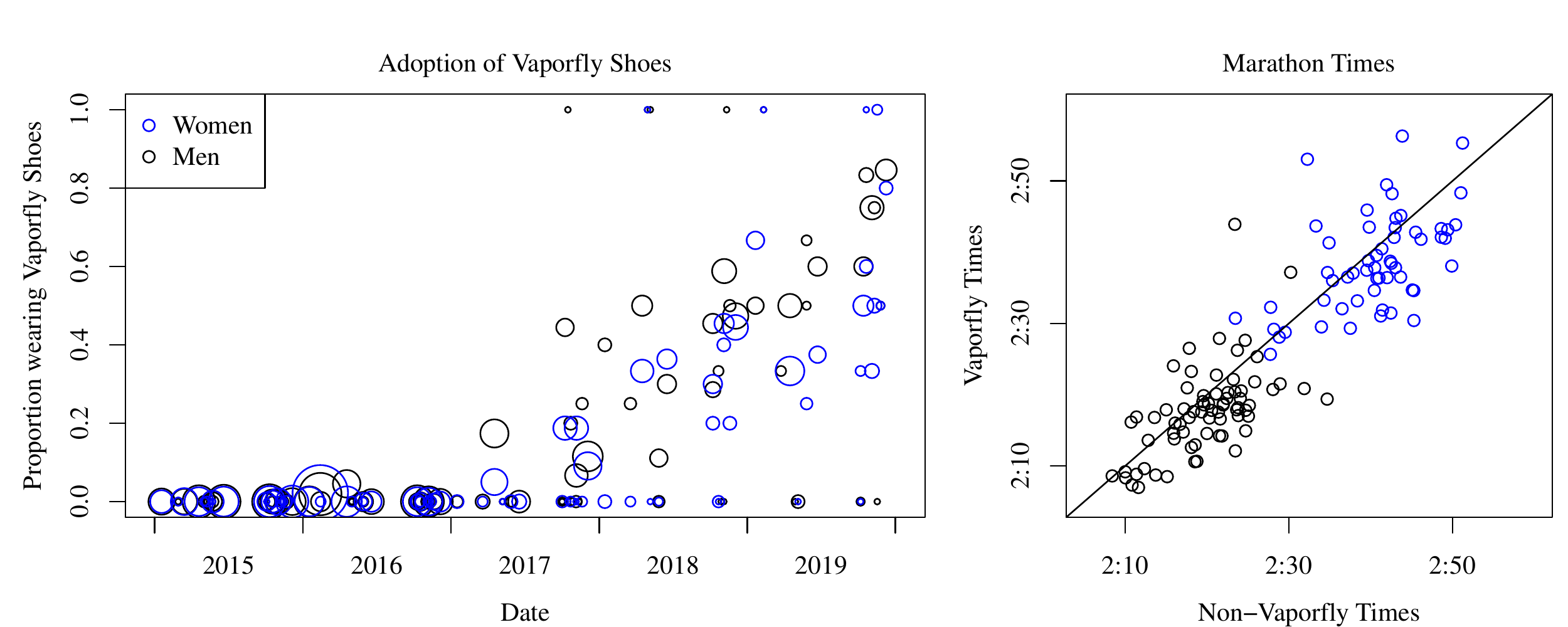}
\caption{\label{data_explore} (Left) Each circle represents
an individual race, with the area of the circle proportional
to the number of runners from the race in our dataset, and the
vertical position equal to the proportion of runners wearing
Vaporfly shoes in the race. (Right) Each circle represents an
athlete, with the horizontal position being the athlete's average
marathon time in non-Vaporfly shoes, and the vertical position
being the athlete's average time in Vaporfly shoes.}
\end{figure}

In Figure \ref{data_explore}, we plot some summaries of the data. The
left plot contains the proportion of runners wearing Vaporflys
in each race in our dataset, separated by sex. Aside from a few
prototypes being used in 2016, adoption of the shoes began in 
early 2017 and rose to over 50\% on average in races at the
end of 2019. The right plot contains the average marathon time
for each athlete in the dataset in Vaporfly vs.\ non-Vaporfly shoes.
Most runners' average time in
Vaporfly shoes is faster than their average time in non-Vaporfly
shoes. Specifically, 53 of 71 men (74.5\%) who switched to
Vaporflys ran faster in them, and 40 of 56 women (71.4\%) who switched
to Vaporflys ran faster in them.

The right plot does not tell the whole story because it might be
the case that runners who switched to Vaporflys did so when
they ran on faster marathon courses. Some courses, such as the
Boston Marathon course, have hills or often have poor weather,
while others are flat and fast. So it is important
to use the data to attempt to account for the difficulty of each course.
%In the right plot, we display the average times at each marathon race
%for all of the athletes in our sampled dataset. We can see that
%there is quite a bit of variability, but this plot is also somewhat
%unsatisfying because we do not know if stronger athletes preferred
%to run at easier or more difficult courses, which would influence
%the averages.
To get a satisfactory estimate of the effect of Vaporfly shoes,
we need to analyze all of the data holistically, controlling
for the strength of each runner and the difficulty of each
marathon course. In the next section, we describe a statistical
model intended for that purpose.

\section{Statistical Model}\label{statistical_model}

We seek to estimate the effect of Vaporfly shoes on marathon
performance, controlling for runner ability and marathon course
difficulty.  This is achieved by fitting a statistical model with
non-random effects for Vaporfly shoes and random effects for
runner ability, overall course difficulty, and course difficulty
specific to year.  To allow for the possibility that men and women
have different performance characteristics, we analyze data from the
two sexes separately. We assign each performance a label between $1$
and $n$ ($n=840$ for men, $n=778$ for women). Each athlete is assigned
a label between $1$ and the number of athletes $A$ ($A=308$ men,
$A=270$ women). We assign each marathon course a label between $1$ and
the number of marathon courses $C$ ($C=22$), and we assign each
individual race a label between $1$ and the number of races $R$,
($R=106$). We summarize our notation for the data here:
\begin{align*}
y_{i} &= \mbox{ marathon time in minutes for performance $i$ }\\
x_{i} &= \left\lbrace \begin{array}{ll}
1 & \mbox{if Vaporfly shoes worn in performance $i$} \\
0 & \mbox{if Vaporfly shoes not worn in performance $i$}
\end{array} \right. \\
j(i) &= \mbox{ label for athlete who completed performance $i$} \\
k(i) &= \mbox{ label for marathon course associated with performance $i$ } \\
\ell(i) &= \mbox{ label for individual marathon race associated with performance $i$} 
%t(i) &= \mbox{ day, since Jan.\ 1 2015, of performance $i$}
\end{align*}

A statistical model is a family of probability distributions that
encodes the assumptions we make about the processes generating the
data. Models generally include unknown parameters relevant to the
questions posed in the study. The goal of the analysis is to use the
data to make inferences about these parameters. We consider the 
following two models for the
performances $y_{1},\ldots,y_n$:
\begin{align*}
\mbox{Untransformed:}& \hspace{-10mm} &Y_{i} &= b_0 + b_1 x_{i} + U_{j(i)} + V_{k(i)} + W_{\ell(i)} + Z_{i} \\
\mbox{Log Untransformed:}& \hspace{-10mm} &\log Y_{i} &=  b_0 + b_1 x_{i} + U_{j(i)} + V_{k(i)} + W_{\ell(i)} + Z_{i} 
\end{align*}
with each of the individual terms defined in the following table
\begin{table}[ht]
\centering
\begin{tabular}{rlll}
Terms & & Assumptions & Description \\
\hline
$b_0, b_1$ & & non-random parameters & $b_1 = $ Vaporfly effect\\
$U_1,\ldots,U_A$ &$\stackrel{ind}{\sim}$& $N(0,\sigma_1^2)$ & runner effects\\
$V_1,\ldots,V_C$ &$\stackrel{ind}{\sim}$ &$N(0,\sigma_2^2)$ & course effects\\
$W_1,\ldots,W_R$ &$\stackrel{ind}{\sim}$ &$N(0,\sigma_3^2)$ & individual race effects\\
$Z_1,\ldots,Z_n$ &$\stackrel{ind}{\sim}$ &$N(0,\sigma_4^2)$ & residual effects
\end{tabular}
\end{table}

\noindent The primary parameter of interest is $b_1$, which is the
effect of the Vaporfly shoes. The model assumes that, all else held
constant, switching to Vaporfly shoes changes the response by
$b_1$. We do not attempt to model Vaporfly effects that vary
among individual runners. The interpretations of the parameters are different
depending on whether we take a log transformation of the times or
not. When modeling untransformed times, the effect of Vaporfly shoes
is additive, meaning that we expect the time to change by adding
$b_1$, and when modeling log-transformed times, the effect is
multiplicative, meaning that we expect the time to change by
multiplying by $\exp(b_1)$.

Aside from $b_0$ and $b_1x_i$, the rest of the terms are independent
normal random effects.  Each of the $A$ runners has its own offset
term $U_j$ to account for the fact that runners have differing
abilities; each of the $C$ marathon courses has its own offset term
$V_k$ to account for the fact that different marathon courses are
slower or faster than others; each of the $R$ individual races has its
own offset term $W_\ell$ to account for the fact that race conditions
vary from year to year, making some years slower or faster than others
at the same course; and finally each of the $n$ individual
performances has a term $Z_i$ to account for any other factors that
affected the performance. We also considered including time-varying
runner effects, to allow each runner's fitness to improve or decline
over time independently of the fitness of other runners, but we
decided that this model placed too much weight on athletes who raced
frequently.  In the Appendix, we also include results from a combined
model for men and women.

\section{Estimated Parameters}\label{estimates}

To fit the models, we used the \verb!lmer! function, which is part of
the \verb!lme4! package \citep{lme4} in the R programming language
\citep{rbib}. The \verb!lme4!  package is a well-established piece of
statistical software for fitting random effects models of the type we
seek to estimate in this research study. Code and data for reproducing
our results are available online at
\url{https://github.com/joeguinness/vaporfly}.

We fit separate models for men and women, and additionally, separate
models for the untransformed and log-transformed marathon times. 
The estimated parameters are summarized in the table below:
\begin{table}[ht]
\centering
\begin{tabular}{crrrr}
 & {men minutes}
 & {women minutes} 
 & {men log minutes}
 & {women log minutes} \\
\cline{2-5}
 & estimate (s.e.) & estimate (s.e.) & estimate (s.e.) & estimate (s.e.) \\
\hline
$b_0$      &   $139.69$ (0.59)  & 159.83 (0.81)       & $4.94$ $(0.004)$ & 5.070 (0.0050) \\
$b_1$      &  $\bm{-2.95}$ $\bm{(0.60)}$ & $\bm{-2.18}$ $\bm{(0.81)}$ & $\bm{-0.0209}$ $\bm{(0.0041)}$ & $\bm{-0.0135}$ $\bm{(0.0049)}$ \\
$\sigma_1$ & 4.175               & 6.40                & 0.030  & 0.041\\
$\sigma_2$ & 1.852               & 2.33                & 0.013  & 0.014\\
$\sigma_3$ & 1.874               & 2.43                & 0.013  & 0.015\\
$\sigma_4$ & 4.108               & 5.02                & 0.028  & 0.030
\end{tabular}
\caption{\label{parameter_estimates} Table of parameter estimates for 
the statistical models.}
\end{table}

For both men and women and for untransformed and log-transformed
times, the Vaporfly effect is negative, indicating that the evidence
supports the hypothesis that Vaporflys decrease, or improve, marathon
times. Our best estimates of the additive effects are $-2.95$ minutes
for men and $-2.18$ minutes for women. Using log-transformed data, our
best estimates of the multiplicative effects are $\exp(-0.0209) =
0.979$ for men, and $\exp(-0.0135) = 0.986$ for women, meaning that we
expect men's times to decrease by $2.06\%$, and women's times to
decrease by $1.34\%$, when wearing Vaporfly shoes, as compared to the
shoes each athlete was wearing before switching to Vaporflys.

While our estimates suggest that the effect of Vaporfly shoes is
greater for men, the estimates come with some uncertainty. In the
following table, we include 90\% confidence intervals for each of
the Vaporfly effects, constructed using a normal approximation to
the sampling distribution of the estimates. 
\begin{table}[h]
\centering
\begin{tabular}{cccc}
 {men minutes} & {women minutes} & {men log minutes} & {women log minutes} \\
\hline
$(-3.933, -1.959)$ & $(-3.514, -0.847)$ & $(-0.028, -0.014)$ & $(-0.022, -0.006)$
\end{tabular}
\caption{\label{confidence_intervals} 90\% confidence intervals for Vaporfly effects in each model.}
\end{table}

\noindent None of the intervals contain zero, which indicates strong evidence
for a non-zero Vaporfly effect. There is substantial overlap between
the men's and women's confidence intervals, which leaves some
uncertainty about which sex benefits most from Vaporfly shoes. In the Appendix,
we include an analysis assessing the difference between the men's and
women's Vaporfly effects; we find that we do not have sufficient 
evidence to conclude that the effects differ by sex.

\begin{figure}
\centering
\includegraphics[width=\textwidth]{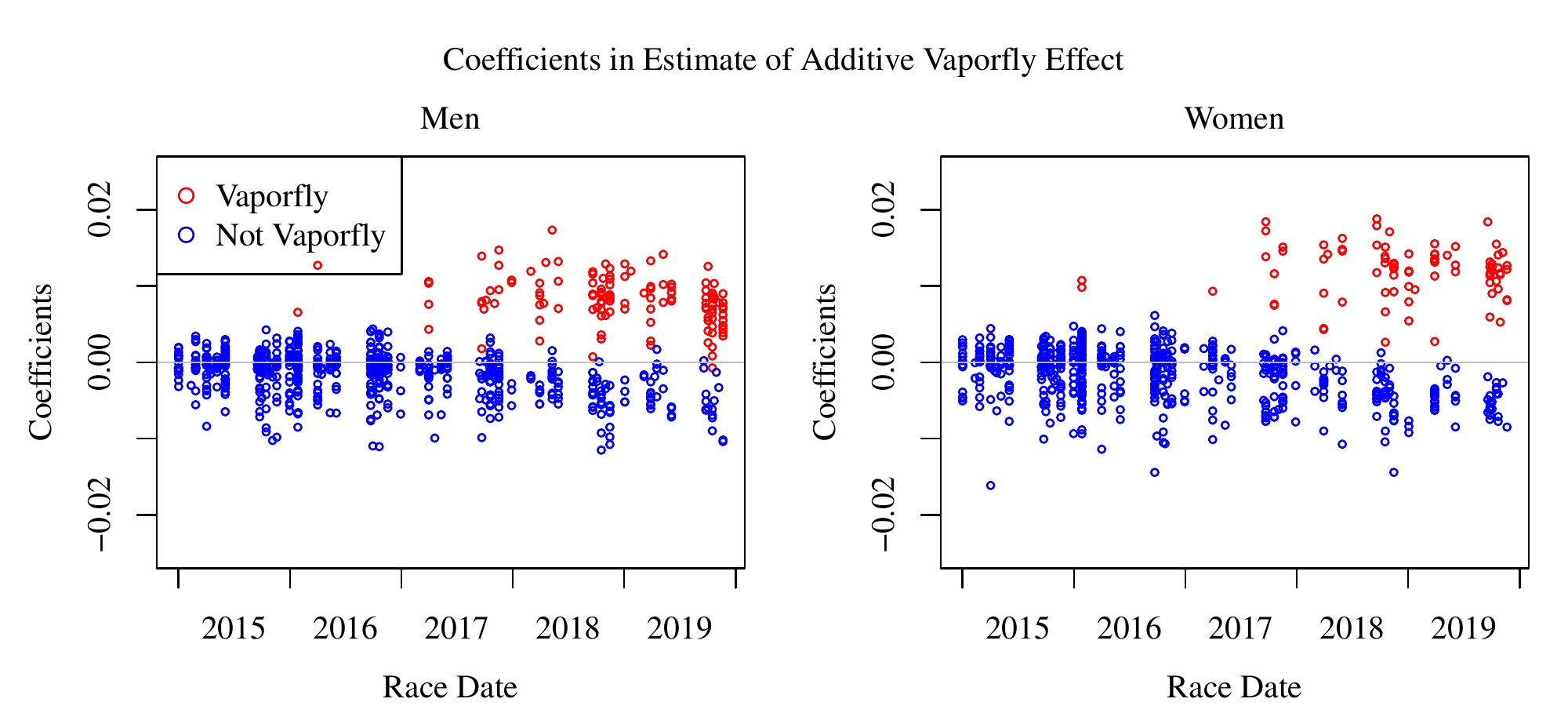}
\caption{\label{vaporfly_coefficients} Coefficients for the
  generalized least squares estimate of the Vaporfly effect
  $\widehat{b}_1$ for the untransformed data. Each point represents an
  individual performance $y_i$. The height of the point is the
  corresponding coefficient $c_i$.}
\end{figure}

In random effects models such as those we use here, the estimates of
the fixed effects, $b_0$ and $b_1$, are calculated using the
generalized least squares criterion. Generalized least squares
attempts to triangulate all of the dependencies in the data, for
example the fact that there are several performances for each runner
and for each race, to arrive at a statistically optimal estimate of
the effects. The estimates are linear combinations of the responses,
for example
\begin{align*}
\widehat{b}_1 = \sum_{i=1}^n c_i y_i,
\end{align*}
where $c_1,\ldots,c_n$ are coefficients calculated using the covariance
matrix of the random effects model. These coefficients can
sometimes have seemingly counter-intuitive values. In the spirit of
attempting to make sense of the magic of generalized least squares,
and to promote its utility for this type of problem, we plot the
coefficients for the estimates of the Vaporfly effects in Figure
\ref{vaporfly_coefficients}.

To help make sense of why generalized least squares picks these
coefficients, consider four performances
\begin{center}
\begin{tabular}{l}
$y_1$ = Time for Runner 1 at Boston Marathon 2016 \\
$y_2$ = Time for Runner 2 at Boston Marathon 2016 \\
$y_3$ = Time for Runner 1 at Chicago Marathon 2017 \\
$y_4$ = Time for Runner 2 at Chicago Marathon 2017 \\
\end{tabular}
\end{center}
The first runner ($y_1$ and $y_3$) did not wear
Vaporflys, but the second runner ($y_2$ and $y_4$) switched to
Vaporflys at Chicago in 2017. A reasonable estimate for the Vaporfly
effect from these data might be the average of the Vaporfly performances
minus the average of the non-Vaporfly performances,
\begin{align*}
y_4 - \frac{1}{3}(y_1 + y_2 + y_3),
\end{align*}
which places a positive coefficient (1.0) on the Vaporfly performance
and negative coefficients ($-0.33$) on the non-Vaporfly performances. However, 
a better estimate would consider how much the second athlete's advantage
increased after switching to the Vaporflys,
\begin{align*}
(y_4 - y_3) - (y_2 - y_1).
\end{align*}
The first difference ($y_4-y_3$) measures how much better the second
athlete did in Chicago (wearing Vaporflys), and the second difference
is how much better the second athlete did in Boston (not wearing
Vaporflys). This estimate places a positive coefficient on the second
runner's Vaporfly performance in Chicago, a positive coefficient on
the first runner's Boston performance, a negative coefficient on the
second runner's Boston performance, and a negative coefficient on the
first runner's Chicago performance.

This pattern can be observed in Figure \ref{vaporfly_coefficients};
early performances---before the Vaporfly appeared on the market---have
either positive or negative coefficients, whereas later performances
generally have positive coefficients when the Vaporfly is worn and
negative coefficients when not worn. There is some variation in
the magnitude of these coefficients, which we expect is due to the
differing number of performances from each runner and from each race.
Note that Figure \ref{vaporfly_coefficients} shows the coefficients
in the generalized least squares estimate, not raw performances.

\section{Discussion}

By collecting data on marathon times and identifying shoes worn in a
systematic sample of elite and sub-elite marathon runners, we studied
how much a runner's marathon time can be expected to improve after
switching to Vaporfly shoes. For men, the improvement is most likely
somewhere between 2.0 and 3.9 minutes, or between 1.4\% and 2.8\%. For
women it is likely between 0.8 and 3.5 minutes, or between
0.6\% and 2.2\%. To put these numbers into perspective, elite marathon 
runners cover more than half a mile in 3 minutes.

We made several assumptions that we believe to be reasonable, but
nevertheless are open for debate and could be refined. We assumed that
the expected Vaporfly effect ($b_1$) is the same for every runner of
the same sex. Including prototypes, there are several different
versions of Vaporflys, and it is logical to expect that newer versions
improve upon older versions. Moreover, depending on biomechanical
factors, some runners may benefit from Vaporflys more than others, or
could have been wearing shoes that were not optimal for them before
switching to Vaporflys. The models assume that, conditional on the
runner and race, the marathon time follows a normal distribution. This
may not be entirely appropriate because we believe that a runner is
more likely to run 5 minutes slower than expected rather than 5
minutes faster; when things go wrong in a marathon, they can go really
wrong.

The study did not include data on runners that did not finish their
races. This data is more difficult to obtain and more difficult to
model. Based on the Wired study \citep{wired1}, which found that people wearing
Vaporflys generally ran better in the second half of the race, we
believe that people wearing Vaporflys are less likely to drop out, so
we are more likely to miss the very worst performances in non-Vaporfly
shoes. Thus, we believe that including drop-outs would only strengthen
our estimate of the Vaporfly effect.

The shoes were identified via a manual process of searching through
photographs. We believe that this manual process for getting the shoe
data is better (though labor intensive) than relying on self-reported
shoes. Nonetheless, it is prone to error in misidentification of the
person and the shoes. For example, Nike has an orange Vaporfly with a
black Swoosh logo, and also a Zoom Fly with the same color scheme.  Another
example, some professional athletes ran in a neon yellow and pink
prototype Vaporfly, which looks very similar to the neon yellow and
pink Nike Zoom Streak 6. See \cite{letsrun1} for a more detailed
analysis. Yet a further example, we identified one athlete who
attempted to conceal the identity of his shoes by coloring in the
white Swoosh on a blue pair of Vaporflys.

Some of the athletes may have competed in marathons not
included in the 22 marathons that we sampled. Missing these
performances shouldn't bias our results, but our results could be
strengthened if we are able to track down every performance from the
athletes in the study.

It is possible that athletes are more likely to switch to 
Vaporfly shoes when they know they are ready to turn in a good
marathon performance. Inversely, some athletes might not be willing
to pay \$250 for shoes when they are out of shape. The Upshot study
investigated this possibility by controlling for training volume; 
they did not see substantially different results. Further, our
sample consists of solely highly accomplished runners. We believe
that these athletes are generally wearing the best shoes available
to them whenever they run a marathon.

We were able to identify shoes in nearly all, but not all marathon
performances.  Athletes have the ability to suppress photographs of
themselves, for example by untagging themselves in Facebook photos, or
simply electing not to post pictures of themselves from their
races. If athletes are more likely to suppress photos of poor
performances in Vaporfly shoes, our estimated effect of Vaporfly shoes
could be larger than it should be.

\vspace{12pt}

\begin{center}
\textbf{\large Acknowledgements}
\end{center}

The authors thank Richard Heffron and Melissa Hardesty for discussions
and providing comments on an early draft of the manuscript, and
Richard Cleary for comments on the first version of the paper. We
thank two peer reviewers, Harry Crane and Ted Westling, for providing
detailed assessments of the paper and suggestions for
improvements. Their reviews, along with our responses, are included in
the Appendix. Finally, we express gratitude for those who take photos
of major marathons and post them to social media, especially Karen
Mitchell, Clay Shaw, and Malrie Sonier.

\bibliography{../refs}{}
\bibliographystyle{apalike}

\appendix

\section{Table of Marathons Used in Analysis}

 \begin{table}[ht]
 \centering
 \begin{tabular}{l}
   \hline
 Boston Marathon \\
   California International Marathon \\
   Chicago Marathon \\
   Columbus Marathon \\
   Eugene Marathon \\
   Grandma's Marathon \\
   Houston Marathon \\
   Indianapolis Monumental Marathon \\
   Los Angeles Marathon \\
   Lakefront Marathon \\
   Marine Corps Marathon \\
   New York City Marathon \\
   Olympic Trials Marathon \\
   Ottawa Marathon \\
   Philadelphia Marathon \\
   Phoenix Marathon \\
   Richmond Marathon \\
   Toronto Waterfront Marathon \\
   Twin Cities Marathon \\
   Vancouver International Marathon \\
   Vermont City Marathon \\
   Wineglass Marathon \\
    \hline
 \end{tabular}
 \end{table}

\section{Peer Review Comments and Responses}

This section is divided into subsections, one for each
reviewer that provided comments on the paper. The reviewers'
comments are italicized, and our responses and discussions
are in normal text.

\subsection{Harry Crane, Rutgers University}

\textit{This is a very interesting analysis of runner performance with and
without the Nike Vaporfly.  I'd expect this to be of interest to the
running and sports science community, and also the general public as
it shows how new technologies can impact the outcome of competitive
events.  In a case such as this, the early adopters will gain a
competitive advantage for however long the benefits of the new product
remain unknown to the broader pool of competitors.  I hope this
analysis will reach the most relevant communities.  I have a few
questions and comments below.  Some of these might be worth commenting
in the paper, but for the most part they are for my own curiosity.}

\textit{1. The analysis notes a performance improvement for athletes who have
switched from non-Vaporfly to Vaporfly.  Are there any instances of
runners going in the other direction, from Vaporfly to non-Vaporfly?
If yes, do you find anything interesting in the first performance back
without Vaporfly?  If not, do the authors believe this is because
there has not been sufficient time to observe such a switch or because
the performance enhancement is noticeable enough to the athletes that
they would not want to switch?}

We identified 5 runners who switched to the Vaporfly and then switched
to a different shoe. Looking closely at each of these cases, we found
that one was a case of a misidentified shoe, bringing the total to 4
runners. One of them joined a racing team sponsored by another company
(Brooks). One switched to what looks like a prototype of Brooks's new
racing shoe. One is wearing a different Brooks shoe, and one is
wearing a different Nike shoe.

This is too small a sample to draw any conclusions, but of these 4
runners, 3 ran faster after switching away from the Vaporflys,
although 1 of the 3 who ran faster did so on a faster course than
the previous marathon in Vaporflys.

It is worth noting that very few athletes switched away from Vaporflys
after trying them.

The analysis has been updated after correcting the misidentified shoe.

\textit{2. Your model includes a random effect for each marathon, which I
suppose captures any factors associated with all runners in that race
running faster or slower than usual, e.g., bad weather, wind, etc.
Have you considered to include a course effect for any marathons in
your dataset that were run over the same course in different years,
e.g., Boston 2016 and Boston 2017.  Would this possibly help tease out
any additional variation in course-day variation?}

This is a good suggestion for making the marathon effects more flexible, 
as you are right that certain courses are difficult every year, 
but the exact difficulty varies from year to year because of weather
conditions and other factors.

We have updated our analysis to include an additional set of random
effects that are specific to marathon courses, in addition to the 
original effects that were specific to course-year combinations. 

Overall, the estimates of the Vaporfly effects decrease slightly in
magnitude, but none of the findings of the paper change
substantially. We think that the decrease in the Vaporfly effects is
due to the added flexibility in the random effects terms.

\textit{3. Presumably athletes who switch to the new shoe would want to know
whether the shoe is going to help their performance specifically,
rather than just a positive association over the population of all
runners who wear the shoe.  Is it possible to draw a causal
relationship from the results of this analysis?  The authors mention
some potential confounding factors, such as cost of the shoe and
runners knowing whether they are training well for an upcoming race.
If one were to attempt such a causal inference, what additional
analyses should be carried out?}

As we understand it, causal inference is an attempt to control for
confounders in an observational study. In that sense, one can interpret
the effect as causal if one believes that all the important confounding
variables have been accounted for. We think we have controlled for
the most important confounders (athlete and marathon) but can't rule out
the possibility that other confounding factors exist, which is why
we mentioned cost and fitness.

Judging by the widespread adoption of the Vaporfly shoes, runners seem
to think that there is a causal effect. Some runners at the 2020 US Olympic 
Trials marathon even wore the new Nike Alphafly shoes after receiving
them a day or two before the race. Conventional running wisdom states
that you should never race in a brand new shoe that you haven't tried
before. In our opinion, this break from conventional wisdom suggests
that runners think the effects are causal.

\textit{4. I'm curious whether any of the authors have actually worn the
Vaporfly. Can the authors provide any anecdotal evidence about their
experience with the shoe, particularly in comparison to their
experience with other (apparently inferior) shoes.}

We haven't worn them yet, though several runners have provided us with
anecdotal evidence that the shoes improve their performances. 

\subsection{Ted Westling, University of Massachusetts Amherst}

\textit{This article presents a clear and well-written study of the
association between wearing the Vaporfly running shoe and race time in
the context of elite marathon runners. I have several comments.}

\textit{1. Why have separate models for women and men? Once you control for
individual runner effects, doesn’t this automatically include sex
differences? I would expect that putting all the runners in a single
model would improve power and also allow a formal assessment of the
interaction of sex and Vaporfly effect via an interaction of the two.}

This is a good question and suggestion. Our main reason for fitting separate models
is that we wanted to avoid assuming that the variances of the random
effects terms were the same for men and women. Looking at the table of parameter
estimates, in the untransformed model, the women's standard deviations
are all higher than the men's standard deviations, which is not surprising
because the women's times are overall larger than the men's. For the log
transformed model, the standard deviations are much closer, with the exception of
the runner standard deviations, which are about 25\% larger for the women.

Thanks to your suggestion, we fit the interaction model and include the results
in the table below, followed by the table of confidence intervals.

\begin{table}[h!]
\centering
\begin{tabular}{crr}
 & {minutes}
 & {log minutes} \\
\cline{2-3}
 & estimate (s.e.) & estimate (s.e.) \\
\hline
$b_0$   &   $140.27$ (0.66)  & 4.94 (0.0043)   \\
$b_1$   &  $-3.35$ $(0.64)$ & $-0.0229$ $0.0041$ \\ 
$b_2$   &  $19.89$ $(0.53)$ & $0.1336$ $0.0035$ \\ 
$b_3$   &  $0.72$ $(0.86)$ & $0.0062$ $0.0055$ \\
$\sigma_1$ & 5.265               & 0.0350         \\
$\sigma_2$ & 2.110               & 0.0137          \\
$\sigma_3$ & 2.239               & 0.0144          \\
$\sigma_4$ & 4.604               & 0.0292          
\end{tabular}
\caption{\label{parameter_estimates_int} Table of parameter estimates for 
the interaction models} 
\end{table}

\begin{table}[h!]
\centering
\begin{tabular}{cccc}
 {men minutes} & {women minutes} & {men log minutes} & {women log minutes} \\
\hline
$(-4.410, -2.298)$ & $(-3.786, -1.478)$ & $(-0.030, -0.016)$ & $(-0.024, -0.009)$
\end{tabular}
\caption{\label{confidence_intervals_int} 90\% confidence intervals
for Vaporfly effects in each model.}
\end{table}

The model is parameterized so that $b_1$ is the men's Vaporfly effect,
and $b_1 + b_3$ is the women's effect, giving men's effects of $-3.35$
and $-0.0229$ for the untransformed and log-transformed models, and
women's effects of $-2.63$ and $-0.0167$. These effects are all larger
in magnitude than those from the sex-separated analyses. We do not
find statistical evidence that the difference between the men's and
women's effects, $b_3$, is different from zero, as the standard error
of the estimates are comparable to the estimates in both models.  For
the untransformed model, the women's confidence intervals get a little
narrower and the men's a little wider, which we suppose is not
surprising because the women's variances were larger in the
non-interaction model. For the log transformed data, the widths of the
confidence intervals do not change much.

\textit{2. As Harry Crane pointed out, controlling for marathon effects doesn’t
account for variation in marathon conditions that vary year-to-year. I
would expect this to substantively impact the conclusions only if
marathon conditions are also associated with the choice to wear
Vaporflys. I would be interested to know whether the authors expect
this to be the case. Relatedly, as part of the data exploration it
would also be interesting to see a plot of the proportion of runners
wearing Vaporflys in each marathon over time.}

This is probably a miscommunication on our part. We do control for
variation from year to year in the same marathon. In the revised
version, we have added effects for each individual course, which do
not vary from year to year. Hopefully, now that we have both types of
effects and describe them separately, the description of the effects
is more clear.

Thank you for the suggestion for the plot. We have replaced our
original plot with one that contains the proportion of runners wearing
Vaporflys in each race as a function of race date.

\textit{3. Regarding the data collection/sampling strategy for the study: were
elite runners first identified, and subsequently all marathon
performances for these runners recorded? Or were marathons first
identified, and only the performances meeting the stated time criteria
included in the study?  In the second strategy, a runner could
theoretically be in the data for one marathon, but excluded for
another due to their time not qualifying. In this case, I’d also be
curious about whether and how the truncation in the sampling strategy
might impact the results.}

We first selected a sample of marathons consisting of 22 of the most
competitive marathons in the United States and Canada. Then from those marathons, we 
selected runners who achieved a minimum performance standard in 2015 or 2016 in
at least one of the 22 selected marathons. Then we selected ALL performances
from those runners, regardless of the time, in those 22 marathons over the 
period 2015 to 2019.

It would probably be better to first create a dataset from all marathons
in 2015 and 2016, and then find runners that met the minimum performance
standard in any marathon in 2015 and 2016. There is some manual work
to be done to get such a dataset, because one needs to know the url
on marathonguide.com for every marathon. We selected the 22 marathons 
to include those where runners had qualified for the U.S.\ Olympic trials,
and marathons that had many Boston Marathon Qualifiers, so we think our 
method misses very few elite runners.
 
\textit{4. The authors refrain from using causal language in the article,
probably because the study is observational rather than experimental
in nature, and making causal conclusions from observational data is
difficult and controversial. However, I don’t think there is any doubt
that the most relevant scientific question is a causal one: for
instance, individual runners might want to know what their time would
be wearing Vaporflys (treatment) versus wearing their usual race shoes
(control). Such an individual-level causal question is challenging to
answer even in the context of randomized trials, but we can and do
strive to answer population-average version of such causal questions
using observational data. The most common approach is to attempt to
adjust for all confounding variables that impact both receipt of
treatment (wearing of Vaporflys) and the outcome (marathon time). It
would be nice if the authors could comment on what variables might be
confounders in this setting.}

We are not experts in causal inference techniques, so we appreciate
your expert opinion on the matter. We list a few possible confounders
in the discussion, namely fitness and price. Since Nike Vaporflys are
expensive relative to other racing shoes, a runner who is in poor
fitness may be unwilling to pay for them to wear in a marathon,
but once a runner gets into good shape, they might be more willing
to pay for the shoes and race in them. Also, some runners might try
the shoes in training and decide that they do not want to race in 
them because they do not get a benefit from them. A runner's personal
objections to the fairness of the shoes might also influence
self-assignment of the shoes.

\end{document}